\documentclass[prd,twocolumn,aps,showpacs,nofootinbib,amsmath,amssymb,floatfix,superscriptaddress,showkeys]{revtex4-1}
\usepackage{multirow}
\usepackage{graphicx}
\usepackage{epstopdf}
\usepackage{dcolumn}
\usepackage{bm}
\usepackage{threeparttable}
\usepackage{subfigure}
\usepackage{color,txfonts}
\usepackage{ulem}
\usepackage{diagbox}
\usepackage{booktabs}
\usepackage{makecell}
\usepackage{array}

\definecolor{blue0}{rgb}{0,0,0.6}
\usepackage[colorlinks,linkcolor=blue0,anchorcolor=blue0,citecolor=blue0,urlcolor=blue0]{hyperref}

\newcommand{\mnras}{Mon. Not. R. Astron. Soc.}

\newcommand{\aap}{Astron. Astrophys}

\newcommand{\apjl}{Astrophys. J.}

\newcommand{\MyTabA}{\ref{MyTabA}}
\newcommand{\MyFigA}{\ref{MyFigA}}
\newcommand{\MyFigB}{\ref{MyFigB}}
\newcommand{\MyFigC}{\ref{MyFigC}}
\newcommand{\MyFigD}{\ref{MyFigD}}
\newcommand{\MyFigE}{\ref{MyFigE}}
\newcommand{\MyFigF}{\ref{MyFigF}}
\UseRawInputEncoding
\begin{document}

\title{Time-dependent stellar-mass binary black hole mergers in AGN disks: Mass \\
distribution of hierarchical mergers}

\author{Guo-Peng Li}
\email[]{lgp@st.gxu.edu.cn}
\affiliation{Laboratory for Relativistic Astrophysics, Department of Physics, Guangxi University,\\
Nanning 530004, China}

\date{\today}

\begin{abstract}
There is much debate about the channels for astrophysical origins of the stellar-mass binary black hole (BBH) mergers
detected by LIGO and Virgo.
Active galactic nuclei (AGNs) are promising sites for the efficient formation and rapid mergers of BBHs
due to migration traps in high-density gas disks within an inner radii.
In this paper, we carry out Monte Carlo simulations to explore  the mass properties
of mergers over time and hierarchical mergers---with one of the black holes (BHs) being the remnant of a previous merger.
We find that the predicted merger rate is $\sim 27-37~{\rm Gpc^{-3}~yr^{-1}}$ and
the detection rate of LIGO/Virgo accompanying with extreme mass-ratio mergers
will increase in the late stage of AGNs.
The fraction of hierarchical mergers is $\sim 24\%$, and its mass-ratio peak is $\sim 0.15-0.35$.
Compared with the low-generation mergers in hierarchical mergers,
the mass ratio of high-generation mergers has a flatter distribution around its peak.
These reveal that the BBH mergers detected by LIGO/Virgo with the extreme mass ratio or heavy component masses
can be well explained in the AGN channel.
\end{abstract}

\maketitle

\section{Introduction}\label{Sec:Intro}
There are more than 80 stellar-mass binary black hole (BBH) mergers detected  by
Advanced LIGO \cite{Aasi_J-2015-Abadie_J-CQG.32.115012} and
Advanced Virgo \cite{Acernese_F-2015-Agathos_M-CQG.32.024001}.
However, there is much debate about the astrophysical formation path to mergers.
The two main channels to form compact binary mergers are
(1) isolated binary evolution \cite{Tutukov_A-1973-Yungelson_L-NInfo.27.70,Dominik_M-2012-Belczynski_K-ApJ.759.52,
Spera_M-2019-Mapelli_M-MNRAS.485.889,van_Son_L_A_C-2021-de_Mink_S_E-arXiv.211001634}
accompanied by mass transfer \cite{van_d_H-2017-E_P_J-MNRAS.471.4256},
common-envelope ejection
\cite{Shao_Y-2021-Li_X_D-arXiv.210703565,Olejak_A-2021-Belczynski_K-A&A.651.A100},
envelope expansion \cite{Tagawa_H-2018-Saitoh_T_R-PhRvL.120.261101},
or chemically homogeneous evolution in a tidally distorted binary \cite{de_M-2016-S_E-MNRAS.460.3545,Marchant_P-2016-Langer_N-A&A.588.A50};
(2) dynamical formation \cite{Mapelli_M-2021-Bouffanais_Y-arXiv.210906222}
in young star clusters
\cite{Banerjee_S-2017-MNRAS.467.524,Rastello_S-2019-Amaro_S-MNRAS.483.1233,Kumamoto_J-2020-Fujii_M_S-MNRAS.495.4268},
globular clusters \cite{Portegies_Z-2000-McMillan_S-ApJ.528.L17,Kamlah_A-2021-Levequear_A-arXiv.210508067},
nuclear star clusters \cite{Antonini_F-2016-Rasio_F-ApJ.831.187,Zevin_M-2019-Samsing_J-ApJ.871.91,
Zhang_F-2019-Shao_L-ApJ.877.87,Fragione_G-2020-Loeb_A-ApJ.902.L26},
or active galactic nucleus (AGN) disks \cite{McKernan_B-2012-Ford_K-MNRAS.425.460,
Bartos_I-2017-Kocsis_B-ApJ.835.165,Yang_Y-2020-Gayathri_V-ApJ.901.L34,McKernan_B-2020-Ford_K-MNRAS.494.1203,
Tagawa_H-2020-Haiman_Z-ApJ.899.26,McKernan_B-2021-Ford_K-arXiv.210707551}.
In addition, there are some alternative scenarios for mergers, such as primordial black hole (PBH) binary mergers \cite{Carr_B_J-1974-Hawling_S_W-MNRAS.168.399,Wang_S-2021-Zhao_Z_C-arXiv.210700450}
and triples or few-body configurations \cite{Antonini_F-2017-Toonen_S-ApJ.841.77,
Liu_B-2019-Lai_D-MNRAS.483.4060}.

Compared with other formation channels, the BBH mergers in the AGN disk have the following properties.
(1) The mass distribution of black holes (BHs) which participate in mergers is heavier than other channels due to the hardening of AGN disks \cite{Yang_Y-2019-Bartos_I-ApJ.876.122}.
This can explain the heavy component masses of BBHs detected by LIGO/Virgo.
(2) Component masses in the ``mass gap'' or a extreme mass ratio become general
via hierarchical BBH mergers (see a review by Ref. \cite{Gerosa_D-2021-Fishbach_M-NatAs.5..749}).
It is shown that in the migration traps,
BHs will merge repeatedly with the remnant of a merger of other two BHs with a high probability
\cite{Bellovary_J_M-2016-Mac_L-ApJ.819.L17,McKernan_B-2018-Ford_K-ApJ.866.66}.
(3) The BBH mergers may accompany with electromagnetic/neutrino signature owing to
a Bondi explosion triggered by hyper-Eddington accretion onto the BBH
\cite{Wang_J_M-2021-Liu_J_R-ApJ.911.L14,Wang_J_M-2021-Liu_J_R-ApJ.916.L17},
ram-pressure stripping of gas within the BH Hill sphere caused by a merger kick
\cite{McKernan_B-2019-Ford_K-ApJ.884.L50},
or high energy neutrinos from choked gamma-ray bursts in AGN disks
\cite{Zhu_J_P-2021-Wang_K-ApJ.911.L19}.

Galactic nuclei are the densest environments of stars and compact objects in the universe \cite{Walcher_C_J-2005-van_d_M-ApJ.618.237}.
In the active phase of galactic nuclei, a high-density gas disk can efficiently assist
binary formation \cite{Bellovary_J_M-2016-Mac_L-ApJ.819.L17,Secunda_A-2019-Bellovary_J-ApJ.878.85, McKernan_B-2020-Ford_K-MNRAS.494.1203}
and mergers \cite{Bartos_I-2017-Kocsis_B-ApJ.835.165}.
There are tens of thousands of stellar-mass black holes harboring within the innermost parsec
in the nuclei of active galaxies due to mass segregation \cite{Morris_M-1993-ApJ.408.496,Hailey_C_J-2018-Mori_K-Natur.556.70}.
These BHs will interact with the AGN accretion disk which helps align their orbits with the AGN disk \cite{Yang_Y-2019-Bartos_I-ApJ.876.122}.
Once the BHs align with the AGN disk, interaction between the gas should move them to migration traps within the disk
\cite{McKernan_B-2012-Ford_K-MNRAS.425.460}.
Generally, the location of migration traps is about 300 Schwarzschild radii from the central supermassive black hole (SMBH)
\cite{Bellovary_J_M-2016-Mac_L-ApJ.819.L17,Secunda_A-2019-Bellovary_J-ApJ.878.85}.
Reference \cite{Peng_P-2021-Chen_X-MNRAS.505.1324} proposed that the existing last migration trap within the AGN disk,
to about a few Schwarzschild radii of SMBH,
such that the mass of a LIGO/Virgo BBH will be higher than the real mass due to doppler and gravitational redshifts.
During the migration to migration traps,
a binary will be formed by the two BHs if the following two conditions are satisfied.
First, the separation of the two BHs is smaller than a mutual Hill radii,
$R_{\rm mH} = [(m_{i} + m_{j}) / 3M_{\rm \bullet}]^{\rm 1/3}[(r_{i} + r_{j})/2]$,
where $m_{i}$ and $m_{j}$ are the masses of the two BHs and $r_{i}$ and $r_{j}$ are their distances
from the SMBH with mass $M_{\rm \bullet}$.
Second, the BBH is hard, i.e., the relative kinetic energy of the BBH,
$k_{\rm rel} = 1/2 \mu v_{\rm rel}^{\rm 2} $,
is less than the binding energy, $U = Gm_{i}m_{j}/(2R_{\rm mH} )$,
where $G$ is the gravitational constant,
and $\mu$ and $v_{\rm rel}$ are the reduced mass of the binary and the relative velocity between the two BHs, respectively.
The binaries are also formed with a higher chance in migration traps \cite{McKernan_B-2020-Ford_K-MNRAS.494.1203}.
Once a BBH is formed, a merger will rapidly occur due to tidal and viscous angular momentum exchange
between the binary and the AGN disk \cite{Bartos_I-2017-Kocsis_B-ApJ.835.165}.

Most of the previous studies focused on migration or migration traps in AGN disks \cite{Bartos_I-2017-Kocsis_B-ApJ.835.165, Secunda_A-2019-Bellovary_J-ApJ.878.85,McKernan_B-2020-Ford_K-MNRAS.494.1203,Tagawa_H-2020-Haiman_Z-ApJ.899.26}.
However, we suggest that orbital alignment of BHs with respect to the disk is more important than migration and migration traps.
Because typical orbital alignment times with the AGN disk are much longer than the time of migration and mergers for BHs \cite{Yang_Y-2019-Bartos_I-PhRvL.123.181101},
and the BHs aligning with the disk would be heavier \cite{Yang_Y-2019-Bartos_I-ApJ.876.122}
and numerous (see more details in Sec. \ref{BPindisk}) in the early stage of AGNs.
These indicate that properties of BBH mergers are different over time.
Therefore, in this paper, we consider time-dependence for mergers by dividing five epochs in the AGN lifetime,
that also can naturally produce a hierarchical merger population.
We investigate mass properties taking into account the effects of the time-dependent and the hierarchical mergers,
and compute the resulting merger masses and mass ratios.

The rest of this paper is organized as follows.
In Sec. \ref{meth} we describe our analytical framework to calculate the formation and mergers of BBHs in AGNs.
In Sec. \ref{R} we present our findings and discuss their implications.
We summarize our conclusions in Sec. \ref{C}.
\section{Method}\label{meth}
We carry out Monte Carlo simulations to obtain the mass distributions of BBH mergers in AGN disks.

\subsection{Stellar-mass black hole population}
We adopt an initial BH mass distribution in the galactic nuclei as
\begin{equation} \label{eq:dN/dm}
\frac{{\rm d}N}{{\rm d}m_{\rm bh}} \propto m^{\rm -\beta}_{\rm bh}.
\end{equation}
The range of BH masses $m_{\rm bh} \in [5M_{\rm \odot},50M_{\rm \odot}]$ is adopted
by considering the lower and pair-instability mass gap.
In our fiducial model, we adopt $\beta = 2.35$ \cite{Yang_Y-2019-Bartos_I-ApJ.876.122}.
It is also indicated that the $\beta-$dependence of results of simulations is weak for $\beta$ in the range $\sim 2 - 3$ \cite{Yang_Y-2019-Bartos_I-ApJ.876.122}.

The mass distribution of BHs changes with the distance from the SMBH
\cite{Bahcall_J_N-1977-Wolf_R_A-ApJ.216.883,Panamarev_T-2019-Just_A-MNRAS.484.3279}.
In general, it is assumed that the semimajor axis $a$ of the orbit of initial BHs from the SMBH obeys
\cite{O'Leary_R_M-2009-Kocsis_B-MNRAS.395.2127,Gondan_L-2018-Kocsis_B-ApJ.860.5}
\begin{equation} \label{eq:dn/da}
\frac{{\rm d}n}{{\rm d}a} \propto a^{-3/2-0.5m_{\rm bh}/m_{\rm max}},
\end{equation}
where $m_{\rm max} = 50~M_\odot$.
The maximal semimajor axis $R_{\rm inf} = 1.2(M_{\rm \bullet} /10^{\rm 6}M_{\rm \odot})^{\rm 1/2} ~{\rm pc}$ \cite{Gultekin_K-2009-Richstone_D_O-ApJ.698.198}
and the minimal semimajor axis $R_{\rm min} \sim 10^{\rm -4} ~{\rm pc}$
\cite{Tagawa_H-2020-Haiman_Z-ApJ.898.25} are adopted.
For simplicity, we assume BHs orbiting the SMBH on circular orbits at inclination angle $\psi$
with $\cos\psi$ being uniform in $[-1,1]$ \cite{Bartos_I-2017-Kocsis_B-ApJ.835.165}.
The number of BHs with the spherical distribution around the SMBH is $\sim(1-4)\times 10^{\rm 4}$ \cite{Generozov_A-2018-Stone_N_C-MNRAS.478.4030}.
We use an initial BH component consisted of $\sim 2\times 10^{\rm 4}$ BHs following
Refs. \cite{McKernan_B-2020-Ford_K-MNRAS.494.1203,Tagawa_H-2020-Haiman_Z-ApJ.898.25}.
We assume that BHs are initially in binaries with a certain fraction $f_{\rm b}$.
Here, we follow Ref. \cite{Tagawa_H-2020-Haiman_Z-ApJ.898.25} to assume $f_{\rm b} = 0.15$ due to soft-binary-single interactions
and deal with the initial binaries using the method of Ref. \cite{Bartos_I-2017-Kocsis_B-ApJ.835.165}.
We adopt the SMBH mass $M_{\rm \bullet} = 10^{\rm 6}~M_{\rm \odot}$ in our simulations.

\subsection{Orbital alignment, migration, and merger}
The BHs that orbit the SMBH at inclination angle $\psi$ with respect to the AGN disk will align with the disk
due to attenuation of their perpendicular velocity $v_{\rm z}$ with respect to the disk.
This is mainly because the BHs will gravitationally capture the gas from the AGN disk every time they cross it.
The mass of the captured gas during each crossing is \cite{Bartos_I-2017-Kocsis_B-ApJ.835.165,Yang_Y-2019-Bartos_I-ApJ.876.122}:
\begin{equation} \label{eq:M_cross}
\Delta M_{\rm cross} = \Delta v\,t_{\rm cross}r^{\rm 2}_{\rm BHL}\pi \Sigma /(2H),
\end{equation}
where $\Delta v$, $t_{\rm cross}$, and $r_{\rm BHL}$
are the relative velocity of the gas and the BH upon crossing,
the duration of the crossing,
and the BH's Bondi-Hoyle-Lyttleton radii, respectively.
With $v_{\rm orb} = (GM_{\rm \bullet}/r)^{\rm 1/2}$ and $t_{\rm orb} = 2\pi r^{\rm 3/2}(GM_{\rm \bullet})^{\rm -1/2}$
being the orbital velocity and the orbital period of the BH with mass $m_{\rm bh}$, one can have
\begin{eqnarray}\begin{array}{c} 
\Delta v = v_{\rm orb}[(1-\rm cos\psi)^{\rm 2}+\rm sin^{\rm 2}\psi]^{\rm 1/2},\\
t_{\rm cross} \approx 2H/(v_{\rm orb}\sin\psi),\\
r_{\rm BHL} = 2Gm_{\rm bh}/(\Delta v^{\rm 2}+c_{\rm s}^{\rm 2}).
\end{array}
\end{eqnarray}
Due to the capture of the gas, the perpendicular velocity $v_{\rm z}$ of the BH will change during each crossing:
\begin{equation}
\frac{\Delta v_{\rm z}}{v_{\rm z}} = \frac{\Delta M_{\rm cross}}{m_{\rm bh}},
\end{equation}
where $v_{\rm z} = v_{\rm orb}\sin\psi$.
Therefore, the timescale of orbital alignment with the disk can be approximated as
\begin{equation}\label{Talign}
\tau _{\rm align} \sim \frac{t_{\rm orb}v_{\rm z}}{2\Delta v_{\rm z}}.
\end{equation}

The BHs will move to the migration traps in the AGN disk analogous to protoplanetary migration due to a torque which is caused by
the exchange of angular momentum between the gas disk and them \cite{McKernan_B-2012-Ford_K-MNRAS.425.460}.
The torque of the BHs will change sign from positive to negative in migration traps, that makes BHs gather together \cite{Bellovary_J_M-2016-Mac_L-ApJ.819.L17,Secunda_A-2019-Bellovary_J-ApJ.878.85}.
It is worthwhile to point out that comparing to the alignment time, the migration time and the merger time are much shorter.
Reference \cite{McKernan_B-2012-Ford_K-MNRAS.425.460} showed that a characteristic time of migration is $10^{\rm 5}~{\rm yr}$.
Reference \cite{Yang_Y-2019-Bartos_I-PhRvL.123.181101} also found that
the migration time of a binary with masses of $30~M_{\rm \odot} -30~M_{\rm \odot}$ is
about $10^{\rm 5}~{\rm yr}$ (see also Ref. \cite{Bartos_I-2017-Kocsis_B-ApJ.835.165}).
In addition,
Ref. \cite{Secunda_A-2019-Bellovary_J-ApJ.878.85} showed that
the merger time is far less than the migration time in the AGN disk.
Therefore, we neglect the time of the migration and mergers in our Monte Carlo simulations.

As a result, BBH formation and mergers between two BHs are equivalent in our simulations \cite{Secunda_A-2019-Bellovary_J-ApJ.878.85,Secunda_A-2020-Bellovary_J-ApJ.903.133}.
In Refs. \cite{Secunda_A-2019-Bellovary_J-ApJ.878.85,Secunda_A-2020-Bellovary_J-ApJ.903.133},
the authors assumed that the gas hardening prescription \cite{Baruteau_C-2011-Cuadra_J-ApJ.726.28}
holds all the way to the GW regime.
This is probably unrealistic, since it is likely that gas hardening becomes less
efficient at small separations (or large binary orbital velocities
around its center of mass) either due to feedback or lower gas mass.
Therefore, it is possible that dynamical hardening is important for
some subsection of the population here, particularly at small separations.
This implies that there could be complicated dynamics at
work (including exchange encounters, binary hardening, or ionization).
In particular, some of the lighter BH-BH binaries could exchange
partners or be ionized in such encounters, and that might change the
results in this paper.

The finding in Ref. \cite{McKernan_B-2020-Ford_K-MNRAS.494.1203} showed
about $80-90\%$ of mergers occur away from migration traps, and $10-20\%$ of mergers occur at traps.
This means that the migration time may be weakly correlated with the mergers because
most of mergers occur within the migration time.
That provides a strong argument for our neglect of the migration time.

\subsection{AGN disk}
We  consider a geometrically thin, optically thick, radiatively efficient, and steady-state accretion disk \cite{Shakur_N-1973-Sunyaev_R_A-A&A.24.337},
which is expected in AGNs \cite{Kocsis_B-2011-Yunes_N-PhRvD.84.024032}.
We assume the viscosity parameter $\alpha = 0.3$ \cite{King_A_R-2007-Pringle_J_E-MNRAS.376.1740}.
We adopt radiation efficiency $\epsilon = L_{\bullet, \rm Edd}/\dot{M}_{\bullet,\rm Edd} c^{\rm 2}= 0.1$ ,
where $L_{\bullet, \rm Edd}$ and $\dot{M}_{\bullet,\rm Edd} $ are the Eddington luminosity and the corresponding accretion rate of
the SMBH, respectively.
Generally, the dimensionless accretion rate $\dot{m}_{\bullet}= \dot{M}_{\bullet}/\dot{M}_{\bullet,\rm Edd}=0.1$
is adopted \cite{Kocsis_B-2011-Yunes_N-PhRvD.84.024032}.
With the given $\alpha$, $\epsilon$, and $\dot{m}_{\bullet}$,
one can estimate the surface density $\Sigma(r)$, the scale height $H(r)$, and the isothermal sound speed $c_{\rm s}(r)$
of the accretion disk around the SMBH based on the Eqs.~(3) - (13) of Ref. \cite{Sirko_E-2003-Goodman_J-MNRAS.341.501}.

It should be noted that the Eq.~(3) of Ref. \cite{Sirko_E-2003-Goodman_J-MNRAS.341.501} is valid only with
$Q\equiv c_{\rm s}\Omega/(\pi G\Sigma)\gtrsim 1$, where $\Omega = ({GM_{\rm \bullet}/r^{\rm 3}})^{\rm 1/2}$ is the Keplerian angular velocity of the disk.
If not, the Eq.~(3) should be replaced by the Eq.~(15) in Ref. \cite{Sirko_E-2003-Goodman_J-MNRAS.341.501}.
In addition, the viscosity is assumed to be proportional to total pressure of the gas,
and the opacity is calculated based on the Eq. (10) of Ref. \cite{Yang_Y-2019-Bartos_I-ApJ.876.122}
(see also Fig.~1 of Ref. \cite{Thompson_A_T-2005-Quataert_E-ApJ.630.167}).
The nature of BH-disk interactions has large uncertainties in inhomogeneous (clumpy) disk beyond
a radii $R_{\rm disk} = 0.1(M_{\rm \bullet} /10^{\rm 6} M_{\rm \odot})^{\rm 1/2} ~{\rm pc}$,
where the AGN disk becomes the self-gravitating with the Toomre Q-parameter, $Q \sim 1$ \cite{Haiman_Z-2009-Kocsis_B-ApJ.700.1952}.
Therefore, we conservatively neglect the BH-disk interactions in this inhomogeneous (clumpy) region of the disk.

\subsection{Monte Carlo simulations}
We divide the AGN lifetime into five epochs, i.e.,
$\{\Delta t_{\rm 1}\in (0,2],~\Delta t_{\rm 2}\in (2,4],~\Delta t_{\rm 3}\in (4,6],~\Delta t_{\rm 4}\in (6,8],~\Delta t_{\rm 5}\in (8,10]\}~{\rm Myr} $.
Here, a fiducial lifetime of $\tau_{\rm AGN} = 10~{\rm Myr}$ is adopted for the AGN
\cite{Yang_Y-2019-Bartos_I-ApJ.876.122,Yang_Y-2019-Bartos_I-PhRvL.123.181101,Yang_Y-2020-Gayathri_V-ApJ.901.L34}.
We assume the formation and mergers will occur by
interaction between the initial single BHs aligning with the AGN disk within the first epoch.
We further assume the initial binary BHs aligning with the disk within the epoch will participate in mergers
but not interact with the other BHs in the disk to form binaries until the next epoch.
Then, the BHs include two parts in the next epoch,
that one is the initial single/binary BHs aligning with the disk in the next epoch
and the other is the remnant BHs coming from the last epoch.
We repeat the above method to calculate mergers in each epoch.

We assume that the binaries are randomly formed within the AGN disk in a certain fraction $f_{\rm bin}$
which is the fraction of BHs in forming BH-BH binaries in the AGN lifetime.
Reference \cite{McKernan_B-2018-Ford_K-ApJ.866.66}  considered the lower limit 0.01 and the upper limit 0.2 for the formation fraction.
But Ref. \cite{Secunda_A-2019-Bellovary_J-ApJ.878.85} showed a larger fraction $\sim 0.6-0.8$ in their simulations.
Reference \cite{Tagawa_H-2020-Haiman_Z-ApJ.898.25} also found the average number of mergers per BH is about 0.4 within 30 Myr.
Therefore, we adopt an appropriate formation fraction with $f_{\rm bin} = 0.5$ in our simulations.

In our simulations, we assume no binary in the AGN disk is disrupted via so called ionization
due to the much shorter merger time than the ionization time \cite{Pfuhl_O-2014-Alexander_T-ApJ.782.101, Bartos_I-2017-Kocsis_B-ApJ.835.165}.
For simplicity, we neglect the increase in mass due to accretion and mass loss via mergers.
We also neglect the merger kick received by the merger remnant
because of  larger orbital velocities $\sim 2\times 10^4~\rm km/s$.
There are four distinct populations of BHs aligning with the disk as follows:
$[a^{\rm +},L^{\rm +}], [a^{\rm +},L^{\rm -}], [a^{\rm -},L^{\rm +}]$ and $[a^{\rm -},L^{\rm -}]$,
where $a^{\rm +}(a^{\rm -})$ and $L^{\rm +}(L^{\rm -})$ represent prograde (reprograde) spins and prograde (reprograde) orbits, respectively \cite{McKernan_B-2018-Ford_K-ApJ.866.66,Wang_Y_H-2021-M_B-arXiv.211003698}.
Here we mainly concentrate on mass distributions in mergers,
and assume the masses of BBH mergers are unaffected on BH spins and orbits.

\section{Results}\label{R}

\subsection{BH populations in AGN disks}\label{BPindisk}
In the AGN lifetime,
we find the fraction $\sim 5.5\%$ of the BHs align with the AGN disk,
in which the fraction of the single BHs is $\sim 5.2\%$ and the fraction of the BHs in binary is $\sim 7.2\%$.
The reason of the different fractions from the single/binary BHs is because
we simulate an initial BBH as a single BH by using its total mass to calculate the alignment time \cite{Bartos_I-2017-Kocsis_B-ApJ.835.165},
while Ref. \cite{Yang_Y-2019-Bartos_I-ApJ.876.122} found heavier BHs align with the disk more quickly.
Our fraction of BHs aligning with the AGN disk is consistent of Ref. \cite{Bartos_I-2017-Kocsis_B-ApJ.835.165}.
However, we find the number of the BHs aligning the disk is different over time in the AGN lifetime.
In our simulations, the $\{43.7\%,18.6\%,14.4\%,12.3\%,11.0\%\}$ of BHs align with the disk within each epoch
$\{\Delta t_{\rm 1},\Delta t_{\rm 2},\Delta t_{\rm 3},\Delta t_{\rm 4},\Delta t_{\rm 5}\}$.
Therefore, it is necessary to calculate mergers as a function of time due to the differences of mergers in masses and numbers.
We consider an epoch with 2~Myr, in which could assure that
there is enough time to interact between most BHs in the AGN disk to form binaries and merge,
but very few BHs have repeat mergers.
For comparison, we also divide the AGN lifetime into four or six epochs and find that our results are robust.
The process (see Sec. \ref{meth}) can naturally comprise a hierarchical BBH merger population to
explore the mergers which have the extreme mass ratio or heavy component masses.

We define that a first generation, or 1g, merger, namely a primary merger,
is the merger of two BHs that each came from the initial BH population.
The merger will be hierarchical if either of the two BHs is the remnant of a pervious merger.
It is worthwhile to be noted that
the previous study (e.g., Ref. \cite{Yang_Y-2019-Bartos_I-PhRvL.123.181101}) only considered a hierarchical merger
in which at least one of the BHs came from the initial BH population.
Here, we suggest that a merger of two BHs
that one is the remnant of a $n$-g merger and the other is the remnant of a $m$-g merger,
is referred to as a $(n+m+1)$-g merger.
If the value of $n$ (or $m$) is 0, the remnant of the $n$ (or $m$)-g merger represents a BH which came from the initial BH population.
It is noted that the interpretation of generation is extended here.
Because, strictly, a merger of two BHs
that one is the remnant of a $n$-g merger and the other is the remnant of a $m$-g merger,
is referred to as a $(\max\{n,m\}+1)$-g merger.
However, if that, properties of mergers between generations may be indistinct.

\begin{table}
\centering
\caption{The fractions of the mergers for different epochs and generations}\label{MyTabA}
\centering{
\begin{tabular}
{l m{1cm}<{\centering}
m{1cm}<{\centering}m{1cm}<{\centering}
m{1cm}<{\centering}m{1cm}<{\centering}
m{1cm}<{\centering}}

\Xhline{0.8pt}\hline\rule{0pt}{12pt}
&\multicolumn{6}{c}{Fraction}\\\Xcline{2-7}{0.8pt}\rule{0pt}{12pt}
&\multicolumn{6}{c}{Epoch}\\\Xcline{2-7}{0.8pt}\rule{0pt}{12pt}
{Generation}& $\Delta t_{\rm 1}$ & $ \Delta t_{\rm 2}$ & $ \Delta t_{\rm 3}$ & $ \Delta t_{\rm 4} $ & $ \Delta t_{\rm 5}$  & All \\\Xhline{0.8pt}\rule{0pt}{8pt}
1g & 21.61$\%$ &15.00$\%$ &14.07$\%$ &13.22$\%$ &12.24$\%$ &76.13$\%$ \\\rule{0pt}{8pt}
2g & 0&2.93$\%$ &4.20$\%$ &5.01$\%$ &5.72$\%$ &17.86$\%$ \\\rule{0pt}{8pt}
3g & 0& 0.23$\%$& 0.83$\%$ &1.47$\%$ &2.09$\%$ &4.62$\%$ \\\rule{0pt}{8pt}
4g & 0 &0&0.11$\%$ &0.33$\%$ &0.64$\%$ &1.08$\%$ \\\rule{0pt}{8pt}
$\ge 5g$& 0& 0 &0.01$\%$ &0.07$\%$& 0.22$\%$ &0.31$\%$ \\
All &21.61$\%$ &18.16$\%$ &19.22$\%$ &20.10$\%$ &20.91$\%$ &100.00$\%$ \\
\hline\hline
\end{tabular}}
\end{table}

In Table~{\MyTabA}, we show the merger fractions in each epoch and each generation.
We find the fractions $\sim 76\%$ and $24\%$ of the primary mergers and the hierarchical mergers, respectively,
that are consistent of the result of Ref. \cite{McKernan_B-2020-Ford_K-MNRAS.494.1203} in bulk mergers
and Ref. \cite{Tagawa_H-2020-Haiman_Z-ApJ.898.25} combining N-body simulations.
Moreover, Ref. \cite{Yang_Y-2019-Bartos_I-PhRvL.123.181101} assumed that the fractions of each generation
obey a Poisson distribution and found that the fraction of hierarchical mergers is up to $\sim 50\%$.
Those suggest that AGN disks are promising sites for the formation and mergers of hierarchical binary BHs.
We see that the fraction of hierarchical mergers increases over time,
particularly among which the fraction of 2g mergers can go up to $\sim 6\%$  in the late stage of the AGN.
In contrast, the fraction of primary mergers reduces over time, but it still dominates the mergers during the AGN lifetime.
In hierarchical mergers, 2g mergers are in the majority and its fraction can reach up to $\sim 18\%$ in the AGN lifetime.
However, the high-generation mergers are smaller than the low-generation mergers,
even exceedingly rare for fifth- or above-generation mergers.
In addition, Table~{\MyTabA} reveals that the fractions of BBH mergers in each epoch are with no difference.
This indicates that the AGN-assisted BBH merger rate is time-independent in the lifetime of the AGN in our simulations.

\subsection{Mass distribution of time-dependent mergers}\label{TVMD}
\begin{figure}
\centering
\includegraphics[width=0.5\textwidth]{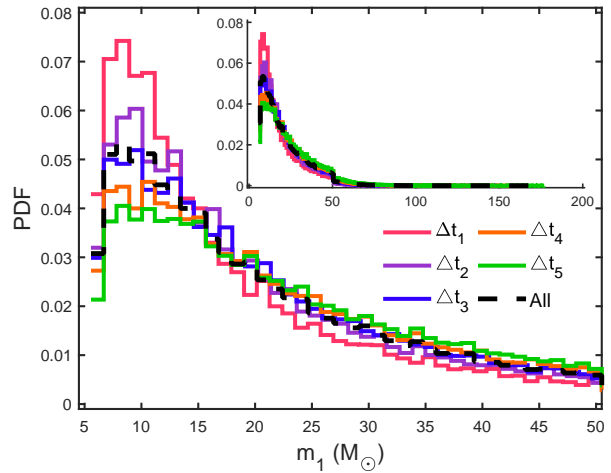}
\caption{The probability density distribution function of primary masses ($m_{\rm 1}$) in mergers for different epochs,
and the distribution function with $m_{\rm 1}$ of all epochs combined.
The main figure shows the distribution function of $m_{\rm 1}$ in the range from $5~M_{\rm \odot}$ to $50~M_{\rm \odot}$,
while the inset shows the distribution function for a wider range of masses.
We show that the epochs
$\Delta t_{\rm 1}\in (0,2]~{\rm Myr}$ (red solid line),
$\Delta t_{\rm 2}\in (2,4]~{\rm Myr}$ (purple solid line),
$\Delta t_{\rm 3}\in (4,6]~{\rm Myr}$ (blue solid line),
$\Delta t_{\rm 4}\in (6,8]~{\rm Myr}$ (orange solid line),
$\Delta t_{\rm 5}\in (8,10]~{\rm Myr}$ (green solid line),
and all epochs with the black dash line.}
\label{MyFigA}
\end{figure}
\begin{figure}
\centering
\includegraphics[width=0.5\textwidth]{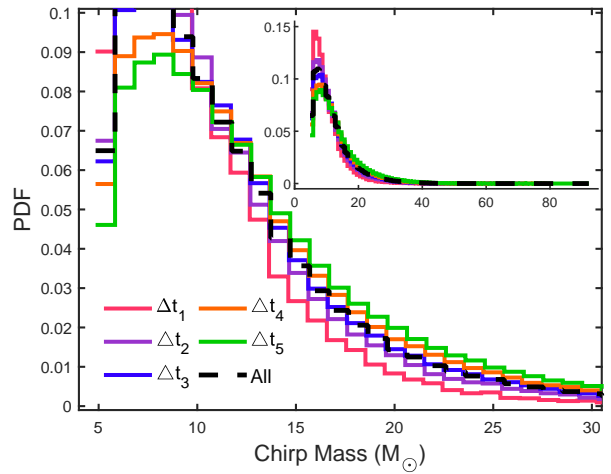}
\caption{The chirp mass ($\mathcal {M}$) distribution function of the mergers in each epoch.
The legend is the same as Fig.~\ref{MyFigA}.}
\label{MyFigC}
\end{figure}
\begin{figure}
\centering
\includegraphics[width=0.5\textwidth]{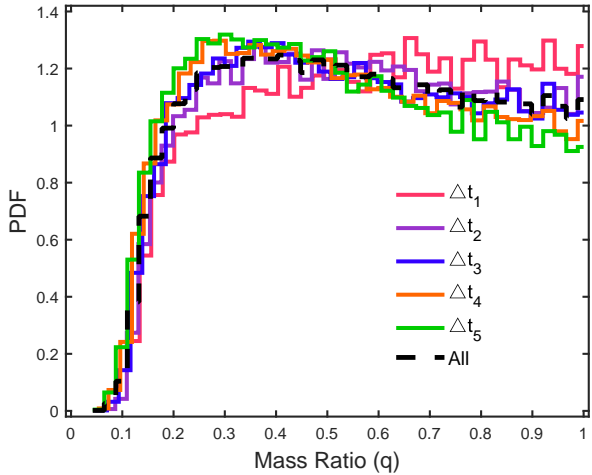}
\caption{The probability density distribution function of mass ratios of BBHs in mergers,
we take account the contribution from each epoch.
The legend is the same as Fig.~\ref{MyFigA}.}
\label{MyFigB}
\end{figure}
In Fig.~{\MyFigA}, we show the primary mass ($m_{\rm 1}$) distribution of mergers in each epoch.
We see that the primary mass distribution increases in the late stage of the AGN
due to the number of hierarchical mergers increases over time.
Therefore, the detection rate of LIGO/Virgo will increase in the late stage of the AGN
because of an invariant merger rate in AGN lifetime but the mass of mergers becomes heavier over time.
This suggests LIGO/Virgo will detect a greater number of mergers in the late stage of the AGN,
in which the fraction of hierarchical mergers  will be expected to exceed $40\%$,
even up to $65\%$ because of the heavier BH masses existing in hierarchical mergers.
To verify the increase of the detection rate, we also plot
the chirp mass ($\mathcal {M} = (m_{\rm 1}m_{\rm 2})^{\rm 3/5}/(m_{\rm 1}+m_{\rm 2})^{\rm 1/5}$) distribution in each epoch in Fig.~{\MyFigC}.

We derive the distribution function of the binaries' mass ratios ($q \equiv m_{\rm 2}/m_{\rm 1}$) for the mergers in different epochs and show it in Fig.~{\MyFigB},
where $m_{\rm 2}$ is the second mass of a BBH and lighter than the primary mass $m_{\rm 1}$.
We find the mass ratio distribution of the epoch $\Delta t_{\rm 1}$ is roughly monotonous and reach the maximum in $q \sim 1$ but
the others have peaks in $q \sim 0.2 - 0.5$.
This indicates that LIGO/Virgo will detect mergers with a extreme mass ratio in the late stage of the AGN.
That may be a possibility for gravitational wave binaries detected by LIGO/Virgo with the extreme mass ratio or heavy component masses,
such as GW170729 \cite{LVC-2019-PhRvX.9.031040}, GW190412 \cite{LVC-2020-PhRvD.102.043015}, or even GW190426$\_$190642 with total mass $\sim 184~M_{\rm \odot}$ reported in GWTC-2.1 \cite{LVC-2021-arXiv.210801045}.
In addition, since Table~{\MyTabA} shows that the fraction of primary mergers still is larger compared to that of hierarchical mergers for each epoch in the AGN lifetime,
the change of the primary mass distribution is not seemingly obvious from the inset of Fig.~{\MyFigA}.
However, it should be noted that the change in the trend will be magnified by LIGO/Virgo
due to extreme mass-ratio mergers in the late stage of the AGN.

\subsection{Mass distribution of hierarchical mergers}\label{HMMD}
\begin{figure}
\centering
\includegraphics[width=0.5\textwidth]{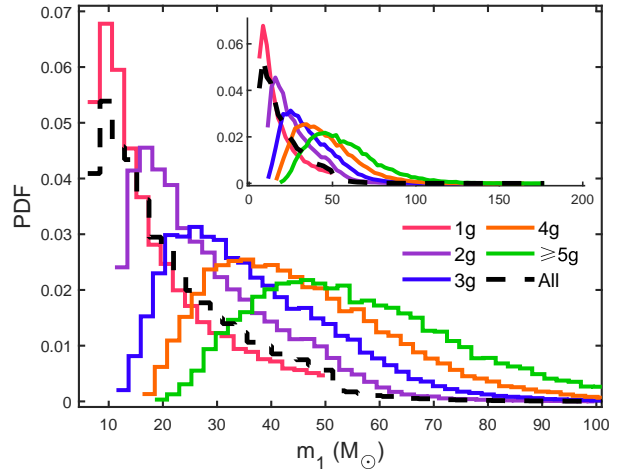}
\caption{The probability density distribution function of primary masses ($m_{\rm 1}$) for different generation mergers,
and the distribution function with $m_{\rm 1}$ of all generations combined.
The main figure shows the distribution function of $m_{\rm 1}$ in the range from $5~M_{\rm \odot}$ to $100~M_{\rm \odot}$,
while the inset shows the distribution function for a wider range of masses.
We show that 1g mergers (red solid line), 2g mergers (purple solid line), 3g mergers (blue solid line), 4g mergers (orange solid line),
5g and above 5g mergers (green solid line), and all mergers with the black dash line. }
\label{MyFigD}
\end{figure}
\begin{figure}
\centering
\includegraphics[width=0.5\textwidth]{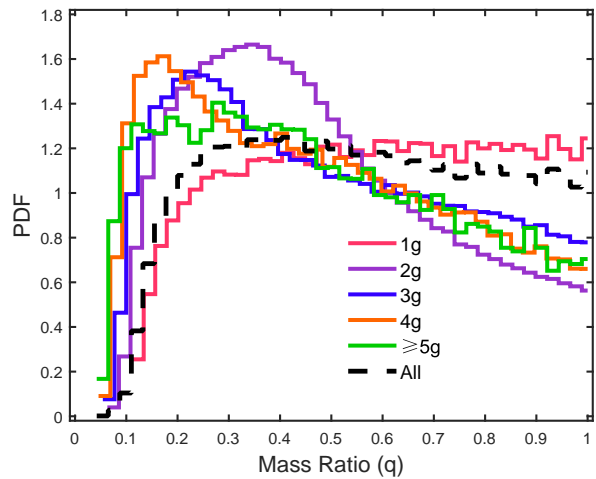}
\caption{The probability density distribution function of mass ratios of BBHs in mergers,
we take account the contribution from each generation.
The legend is the same as Fig.~\ref{MyFigD}.}
\label{MyFigE}
\end{figure}

In Fig.~{\MyFigD}, we show the primary mass distribution for different generation mergers.
We see that the primary mass distribution increases significantly in high-generation mergers as expected,
that is same as the distribution for different epochs.
Reference \cite{Yang_Y-2019-Bartos_I-PhRvL.123.181101} showed that a fraction $\sim 30\%$ of mergers
will have a BH with mass higher than
the $50~M_{\odot}$ upper limit expected from stellar evolution
\cite{Belczynski_K-2016-Heger_A-A&A.594.A97, Woosley_A_E-2017-ApJ.836.244,Giacobbo_N-2018-Mapelli_M-MNRAS.480.2011}.
However, our results cannot reach the high percentage
because of about $24\%$ of hierarchical mergers is obtained in our simulations
but about $50\%$ is found in their results.

\begin{figure*}
\centering
\begin{minipage}{0.49\linewidth}
\includegraphics[height=0.6\textwidth, width=0.9\textwidth]{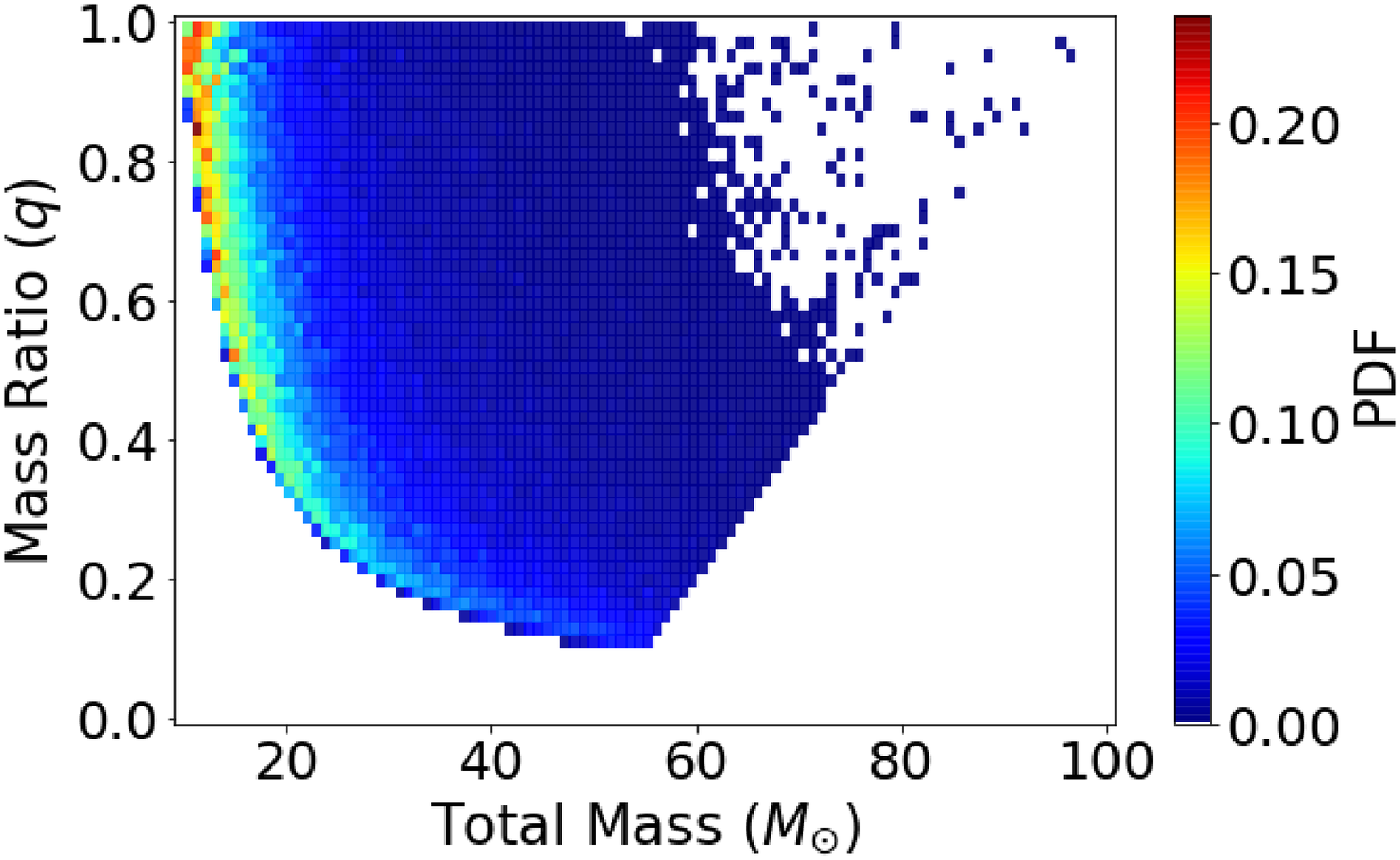}
\end{minipage}
\begin{minipage}{0.49\linewidth}
\includegraphics[height=0.6\textwidth, width=0.9\textwidth]{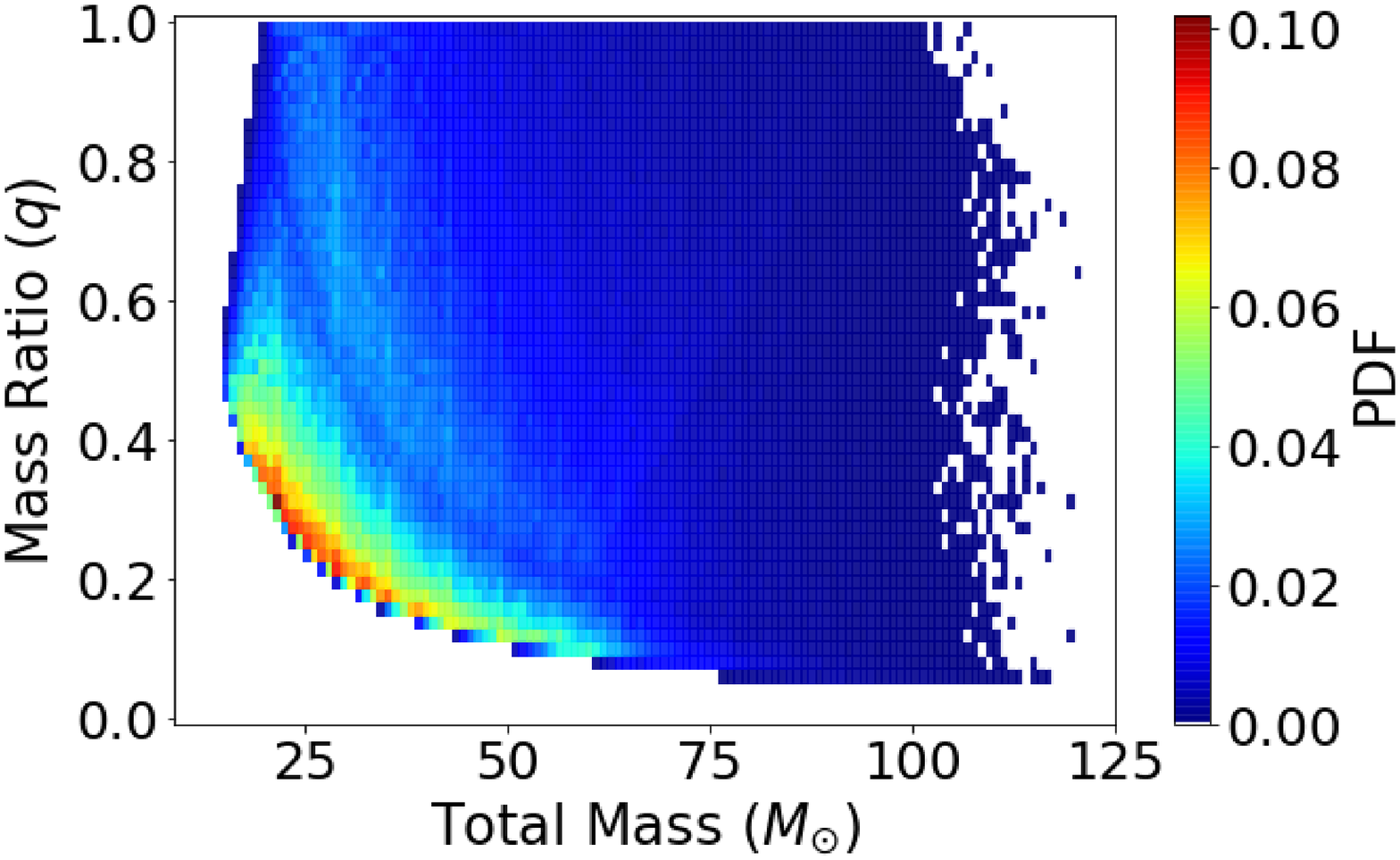}
\end{minipage}
\caption{2D probability densities of the total mass and the mass ratio for the primary (left) and hierarchical (right) mergers.}
\label{MyFigF}
\end{figure*}

We calculate the probability density distribution of mass ratios of the binaries for different generation mergers
and show it in Fig.~{\MyFigE}.
We see that the mass ratio of the hierarchical mergers has the peak in the range about from 0.15 to 0.35,
in which the peak of 4g mergers can reach to $\sim 0.15$.
Reference \cite{Fishbach_M-2020-Tarr_W_M-ApJ.891.L31} predicted that $99\%$ of events should have $q\gtrsim 0.5$
via the BH population detected during the first and second observing run of Advanced LIGO (O1 and O2).
However, we find that about $45\%$ of mergers can have $q < 0.5$ in the AGN disk,
and it can rise up to $60\%$ for the hierarchical mergers.
Furthermore, it is worth to be noted that the mass ratio of $\ge$ 5g mergers distinctly
has an approximate flat distribution around its peak,
and the value of the peak is also smaller than other hierarchical mergers.
Because we calculate hierarchical mergers including the mergers
which are the results of merging from two BHs that at least each came from the remnant of a 1g merger.
That will induce the decline of the value of $q$ around its peak in high-generation mergers.
This also is a reason that the value of the peak is larger in second-generation mergers but not third- or above-generation mergers.
Therefore, we suggest that the mass ratio of high-generation mergers has a flatter distribution around its peak
than low-generation mergers in hierarchical mergers.
Anyway, the minimum of $q\sim 0.03$ still comes from highest-generation mergers, i.e., $\ge$ 5g mergers.
The extreme unequal masses of mergers detected by LIGO/Virgo will be well explained if these occurred in AGN disks.
We also show 2D probability densities of the total mass and mass ratio for mergers in the AGN disk in Fig.~{\MyFigF}.
In the left panel, we plot mergers only including the primary mergers, and in the right is the hierarchical mergers.

\subsection{Merger rate}\label{MR}
Reference \cite{McKernan_B-2018-Ford_K-ApJ.866.66} parameterized the rate of BH-BH mergers in AGN disks simply as:
\begin{align}
R = & 12~{\rm Gpc^{-3}~yr^{-1}}
\frac {N_{\rm GN}} {0.006~{\rm Mpc^{-3}}}
\frac {N_{\rm BH}} {2\times10^4}
\frac {f_{\rm AGN}} {0.1} \nonumber\\
& \times \frac{f_{\rm d}} {0.1}
\frac{f_{\rm bin}} {0.1}
\frac{\epsilon} {1}
\left( \frac {\tau_{\rm AGN}} {10~{\rm Myr}} \right) ^{-1},
\end{align}
where $N_{\rm GN}$ is the average number density of galactic nuclei in the universe,
$N_{\rm BH}$ is the number of BHs in an AGN disk,
$f_{\rm AGN}$ is the fraction of galactic nuclei that
have active AGNs that last for time $\tau_{\rm AGN}$,
$f_{\rm d}$ is the fraction of BHs that end up in the AGN disk,
and $\epsilon$ represents the fractional change in $N_{\rm BH}$ over one full AGN duty cycle.
Using our finding that, within $1.2~{\rm pc}$ of an AGN disk $5.5\%$ of BHs end up in the disk,
one can give a merger rate of $33~{\rm Gpc^{-3}~yr^{-1}}$.

We noticed that the maximal semimajor axis $R_{\rm inf}$ of the AGN disk used here is $1.2~{\rm pc}$,
which may be very large $O(10^7)~r_{\rm g}$ for a $10^6~M_{\rm \odot}$ SMBH.
Even if we are only considering BH migration within $0.1~{\rm pc}$ due to $Q<1$ outside this radii,
that is $O(10^6)~r_{\rm g}$ which is still very large.
Note that migration time which can be short in the inner disk depends on the semimajor axis and surface density of the disk.
Type I migration timescale will be very long in the outskirts of the disk model, i.e., in a Ref. \cite{Sirko_E-2003-Goodman_J-MNRAS.341.501} model disk type I migration timescale is $>10~{\rm Myr}$ for objects of $10~M_{\rm \odot}$
at the radii of the disk $>10^4~r_{\rm g}$.
Most of the initially embedded population will be at large
radii and not make it into the inner disk (near migration traps).
However, any mergers that occur far out in the inner disk require very efficient gas hardening,
or a random initially hard binary.

Therefore, we check how the rate of mergers change by making the AGN disk radii to $10^5~r_{\rm g}$ or $10^4~r_{\rm g}$.
We find that within $10^5~r_{\rm g}$ and $10^4~r_{\rm g}$ of an AGN disk $\sim 6.2\%$ and $4.5\%$ of BHs align with the disk
in the AGN lifetime, respectively.
The difference in the fractions depends largely on the surface density of the disk which has the maximum
between $10^4~r_{\rm g}$ and $10^5~r_{\rm g}$ ($\tau _{\rm align} \propto \Sigma^{-1}$).
Based on these, we find the allowed range of the BBH merger rate is
$R\sim 27-37~{\rm Gpc^{-3}~yr^{-1}}$.

\section{Conclusions}\label{C}
In this paper, we examine mass distributions of BBH mergers in AGN disks
and consider the differences in the time-dependent mergers and hierarchical mergers.
Note that according to Eq.~(\ref{Talign}), in fact we neglect
the change of radii of BHs when these align with
the disk (i.e., ${\rm d}a/{\rm d}t \approx 0$ per orbit)
and only concentrate on the timescale of orbital alignment.
Reference \cite{Secunda_A-2019-Bellovary_J-ApJ.878.85} found that the last BHs to reach the migration
trap region are the innermost BHs which have a large initial semimajor axes and
small inclination angle \cite{Fabj_G-2020-Nasim_S_S-MNRAS.499.2608}.
Therefore, time (not migration time) from BHs
aligning the disk to formation (mergers) of binaries is strongly dependent on
radii of BHs when these align with the disk. This may largely impact on the
binaries¡¯ effective spin
($\chi_{\rm eff} = (m_1 \chi_1 \cos \theta_1+m_2 \chi_2 \cos \theta_2)/(m_1+m_2)$,
where $\chi_i$ and $m_i$ are the spin and mass, respectively, of each
BH in the merged BBH, and the spin obliquity $\theta_i$ is the
angle between the spin $\chi_i$ vector and the orbital angular momentum vector)
but not mass and mass ratio because we do not consider the
migration time which is much shorter than the alignment time.

For $\chi_{\rm{eff}}$, it should have a large value by merging due to
gas accretion from the AGN disk that will tend to torque the BH spin
into alignment with the gas after about 1-10\% of the gas has been
accreted (e.g., Ref. \cite{Bogdanovic_T-2007-Reynolds_C-ApJ.661.147}).
After $10~{\rm Myr}$ any BHs that have
been present in the disk should have been torqued into alignment with
the disk orbital angular momentum.
Therefore, the population of mergers in the late AGN disk (and associated mass and mass ratio
distributions) should therefore be biased towards large values of $\chi_{\rm{eff}}$.
Early populations are more likely to have
random spin alignments and therefore a range of $\chi_{\rm{eff}}$ centered around
zero. So, by testing the population of early and late AGN disks,
we also could probe the likely torquing by disk accretion onto the
embedded objects.

Our conclusions are the following:

(1) The BBH merger rate which is time-independent (see Table~{\MyTabA}) is $\sim 27-37~{\rm Gpc^{-3}~yr^{-1}}$ in the AGN lifetime.
However, the detection rate of LIGO/Virgo
accompanying with extreme mass-ratio mergers (see Fig.~{\MyFigB})
will increase over time (see Fig.~{\MyFigC})
due to heavy component masses (see Fig.~{\MyFigA}) in the AGN lifetime.

(2) The fraction of the hierarchical mergers is $\sim 24\%$, in which 2g mergers dominate with $\sim 18\%$ (see Table~{\MyTabA}).

(3) The peak of the mass ratio of hierarchical mergers is $q \sim 0.15-0.35$, in which 4g can reach to $\sim 0.15$
(see Fig.~{\MyFigE}, also Fig.~{\MyFigF}).

(4) The mass ratio of the high-generation mergers has an approximate flatter distribution around its peak
than the low-generation mergers (see Fig.~{\MyFigE}), and 2g mergers have larger value of the peak in mass ratios.
Because a hierarchical merger of two BHs that each possibly came from the remnant of a previous merger.
Such consideration is different from Ref. \cite{Yang_Y-2019-Bartos_I-PhRvL.123.181101}.

(5) About $45\%$  of BBH mergers in AGN disks have mass ratio $q<0.5$ (see Fig.~{\MyFigB}), while it can go up to $\sim 60\%$ for hierarchical mergers (see Fig.~{\MyFigE}).

(6) If the BBH mergers detected by LIGO/Virgo with the extreme mass ratio
or heavy component masses can be well explained in the AGN channel.


\begin{acknowledgments}
We thank the anonymous referee for beneficial suggestions that improved the paper.
We are grateful to Da-Bin Lin for useful suggestions.
This work is supported by
the National Natural Science Foundation of China (Grant No. 11773007) and
the Guangxi Science Foundation (Grant No. 2018GXNSFFA281010).
\end{acknowledgments}

%

\end{document}